\documentclass[english]{llncs}
\usepackage[latin9]{inputenc}
\usepackage{babel}
\usepackage{url}
\usepackage{graphicx}
\usepackage[unicode=true,pdfusetitle,
 bookmarks=true,bookmarksnumbered=false,bookmarksopen=false,
 breaklinks=true,pdfborder={0 0 0},backref=false,colorlinks=false]
 {hyperref}

\makeatletter
\usepackage{import}

\usepackage{ifthen}
\usepackage{url}
\usepackage{amssymb}
\usepackage{amsmath}
\usepackage{import}
\usepackage{xparse}
\usepackage{amsmath}
\usepackage[usenames,dvipsnames]{xcolor}
\usepackage{xspace}
\usepackage{datatool}
\usepackage{glossaries}

\newcommand{\m}[1]{\ensuremath{#1}\xspace}

	\newcommand{\lrule}{\m{\leftarrow}}
	\newcommand{\cause}{\m{\stackrel{c}{\lrule}}}

	\newcommand{\struct}{\m{I}}

	\NewDocumentCommand\inter{g+g}{%
	  \IfNoValueTF{#1}
	    {\struct}
	    {\m{#1^{#2}}}}

	\newcommand{\bool}{\m{\mathbb{B}}}

	\renewcommand{\int}{\m{\mathbb{Z}}}

	\NewDocumentCommand\subs{g+g}{%
	  \IfNoValueTF{#1}
	    {\m{/}}
	    {\m{#1/ #2}}}

	\newcommand{\logicname}[1]{\text{\sc #1}\xspace}

	\newcommand{\fodot}{\logicname{FO(\ensuremath{\cdot})}}

\newcommand{\ouracronym}[3]{%
	\newacronym{#1}{#2}{#3}
	\expandafter\newcommand\csname #1\endcsname{\gls{#1}\xspace}%
}
	\ouracronym{FO}{FO}{first-order logic}
	\ouracronym{MX}{MX}{Model Expansion}
	\ouracronym{MO}{MO}{Model Optimization}
	\ouracronym{ASP}{ASP}{Answer Set Programming}
	\ouracronym{TP}{TP}{Theorem Proving}
	\ouracronym{LP}{LP}{Logic Programming}
	\ouracronym{CP}{CP}{Constraint Programming}
	\ouracronym{FP}{FP}{Functional Programming}
	\ouracronym{KR}{KR}{Knowledge Representation}
	\ouracronym{CSP}{CSP}{Constraint Satisfaction Problem}
	\ouracronym{SMT}{SMT}{SAT Modulo Theories}
	\ouracronym{KBS}{KBS}{Knowledge Base System}
	\ouracronym{NNF}{NNF}{Negation Normal Form}
	\ouracronym{FNNF}{FNNF}{Flat Negation Normal Form}
	\ouracronym{DefNNF}{DefNNF}{Definition Negation Normal Form}
	\ouracronym{CDCL}{CDCL}{Conflict-Driven Clause-Learning}
	\ouracronym{LCG}{LCG}{Lazy Clause Generation}

	\makeatletter
	\def\ifenv#1{
	\def\@tempa{#1}%
	\def\@ttempa{#1*}%
	\ifx\@tempa\@currenvir
	\expandafter\@firstoftwo
	\else
	\expandafter\@secondoftwo
	\fi
	}
	\makeatother

	\newcommand{\ddrule}[4]{\ensuremath{#1 \leftarrow #2 & \{#3\} & #4}}
	\newcommand{\drule}[2]{\ensuremath{#1 & \leftarrow & #2}}

	\newcommand{\darule}[4]{\ensuremath{#1 \leftarrow #2 & \{#3\} & #4}}
	\newcommand{\arule}[2]{\ensuremath{#1 \, &\leftarrow \, #2}}

	\newcommand{\LNDRule}[2]{
	\ifenv{array}
	{\drule{#1}{#2}}
	{ \ifenv{align}
		{\arule{#1}{#2}}
		{\ifenv{align*}
		{\arule{#1}{#2}}
		{ERROR: using LDRule in unsupported environment: \@currenvir}
		}
	}
	}

	\newcommand{\LDRule}[4]{
	\ifenv{array}
	{\ddrule{#1}{#2}{#3}{#4}}
	{ \ifenv{align}
		{\darule{#1}{#2}{#3}{#4}}
		{\ifenv{align*}
		{\darule{#1}{#2}{#3}{#4}}
		{ERROR: using LDRule in unsupported environment: \@currenvir}
		}
	}
	}

	\NewDocumentCommand\LRule{m+g+g+g}{%
		\IfNoValueTF{#2}%
		{#1.&}{%
		\IfNoValueTF{#3}
		{\LNDRule{#1}{#2.}}
		{\LDRule{#1}{#2.}{#3}{#4}}%
		}
	}

	\NewDocumentCommand\CLRule{m+g}{%
	\ifenv{array}
	{\cdrule{#1}{#2}}
	{ \ifenv{align}
		{\carule{#1}{#2}}
		{\ifenv{align*}
			{\carule{#1}{#2}}
			{ERROR: using CLRule in unsupported environment: \@currenvir}
		}
	}
	}

	\NewDocumentCommand\carule{m+g}{%
		\IfNoValueTF{#2}
			{\ensuremath{#1.}}
			{\ensuremath{#1 \, &\cause \, #2}}}
	\NewDocumentCommand\cdrule{m+g}{%
		\IfNoValueTF{#2}
			{\ensuremath{#1.}}
			{\ensuremath{#1 & \cause & #2}}}

	\newcommand{\algrule}[4]{
	\hbox{{#1}:}& 
	\quad #2 ~\longrightarrow~ #3 
	\hbox{~ if } #4\\
	}

	\newcommand{\AlgoRule}[4]{
	\ifenv{array}
	{\algrule{#1}{#2}{#3}{#4}}
		{ERROR: using AlgoRule in unsupported environment: \@currenvir}
	}

\newcommand{\commentstyle}{\color{Gray}}

	\usepackage[final]{listings}

	\lstdefinelanguage{idp}{
		morekeywords=[1]{namespace,vocabulary,theory,structure,procedure,term,set,formula, spec, specification},
		morekeywords=[2]{include,using,type,isa,contains,partial,extern,LFD,GFD,constructed,from,constraint,func,pred,supertype,of,subtype,define},
		morekeywords=[3]{int,float,char,string,nat},
		morekeywords=[4]{if,then,else,for,end},
		morecomment=[s]{/*}{*/},	
		morecomment=[l]{//}
	}
	\newcommand{\ignore}[1]{}

	\newboolean{nocomments}
	\setboolean{nocomments}{false}

	\newboolean{commentmargin}
	\setboolean{commentmargin}{true}

	\newcommand{\namedcomment}[3]{
		\ifthenelse{\boolean{nocomments}}
		{} 
		{ 
			\ifthenelse{\boolean{commentmargin}}
				{ {\color{#3} \marginpar{\color{#3}\sc #2}#1}  } 
				{  {\color{#3} {\sc #2}: #1}  } 
		}
	}
	\newcommand{\mnamedcomment}[3]{\ifthenelse{\boolean{nocomments}}{}{{\marginpar{ \color{#3}{\sc #2}:#1}}}}

	\usepackage{soul}

\usepackage{etoolbox}

\newcommand\setcitation[2]{%
  \csdef{mycommoncitation#1}{#2}}

\setcitation{IDP}{WarrenBook/DeCatBBD14}
\setcitation{fodot}{tocl/DeneckerT08}
\setcitation{CP}{Apt03}
\setcitation{EZCSP}{lpnmr/Balduccini11}
\setcitation{KR}{Baral:2003}
\setcitation{ASPComp2}{lpnmr/DeneckerVBGT09}
\setcitation{ASPComp3}{journals/tplp/CalimeriIR14}
\setcitation{ASPComp4}{conf/lpnmr/AlvianoCCDDIKKOPPRRSSSWX13}
\setcitation{CPSupport}{ictai/DeCat13}
\setcitation{functionDetection}{iclp/DeCatB13}
\setcitation{fodot2asp}{DeneckerLTV12} 
\setcitation{Tarskian}{DeneckerLTV12} 
\setcitation{Inca}{iclp/DrescherW12}
\setcitation{csp2asp}{ijcai/DrescherW11}
\setcitation{DPLLT}{cav/GanzingerHNOT04}
\setcitation{AspInPractice}{synthesis/2012Gebser}
\setcitation{clasp}{ai/GebserKS12}
\setcitation{oclingo}{kr/GebserGKOSS12}
\setcitation{gringo}{lpnmr/GebserST07}
\setcitation{cmodels}{aaai/GiunchigliaLM04}
\setcitation{inputster}{tplp/Jansen13}
\setcitation{DLV}{tocl/LeonePFEGPS06}
\setcitation{LearningPaper}{TPLP/BruynoogheBBDDJLRDV} 
\setcitation{clog}{iclp/BogaertsVDV14} 
\setcitation{foc}{iclp/BogaertsVDV14}
\setcitation{inferenceClog}{ecai/BogaertsVDV14}
\setcitation{examplesClog}{nmr/BogaertsVDV14b} 
\setcitation{AFT}{DeneckerBV12}
\setcitation{KBS}{iclp/DeneckerV08}
\setcitation{KBPE}{inap/DePooterWD11}
\setcitation{lazyGrounding}{corr/CatDSB14} 
\setcitation{ASP}{marek99stable}
\setcitation{satid}{sat/MarienWDB08}
\setcitation{lazyclausegeneration}{constraints/OhrimenkoSC09}
\setcitation{FP}{ACMCS/Hudak89}
\setcitation{GroundingWithBounds}{jair/WittocxMD10}
\setcitation{SAT}{faia/SilvaLM09}
\setcitation{LTC}{iclp/Bogaerts14}
\setcitation{SPSAT}{ictai/DevriendtBMDD12}
\setcitation{BreakID}{cspsat/DevriendtBB14}
\setcitation{LCG}{stuckeyLCG}
\setcitation{MiniZinc}{conf/cp/NethercoteSBBDT07}
\setcitation{amadini}{cpaior/AmadiniGM13}
\setcitation{bootstrapping}{lash/BogaertsJDJBD14}
\setcitation{GroundedFixpoints}{ai/BogaertsVD15}
\setcitation{LogicBlox}{datalog/GreenAK12}
\setcitation{proB}{journals/sttt/LeuschelB08}
\setcitation{NaturalInductions}{KR/DeneckerV14} 
\setcitation{LP}{jacm/EmdenK76}
\setcitation{SMT}{faia/BarrettSST09}
\setcitation{AF}{ai/Dung95}
\setcitation{ADF}{kr/BrewkaW10}
\setcitation{ADFRevisited}{ijcai/BrewkaSEWW13}
\setcitation{DefaultLogic}{ai/Reiter80}
\setcitation{DL}{ai/Reiter80}
\setcitation{AEL}{mo85}
\setcitation{}{}
\setcitation{completion}{adbt/Clark78}
\setcitation{}{}
\setcitation{}{}
\setcitation{}{}
\setcitation{}{}
\setcitation{}{}
\setcitation{}{}
\setcitation{}{}
\setcitation{}{}
\setcitation{}{}
\setcitation{}{}
\setcitation{}{}
\setcitation{}{}
\setcitation{}{}
\setcitation{}{}
\setcitation{}{}
\setcitation{}{}
\setcitation{}{}
\setcitation{}{}
\setcitation{}{}
\setcitation{}{}
\setcitation{}{}
\setcitation{}{}
\setcitation{}{}
\setcitation{}{}
\setcitation{}{}
\setcitation{}{}
\setcitation{}{}
\setcitation{}{}
\setcitation{}{}
\setcitation{}{}
\setcitation{}{}
\setcitation{}{}
\setcitation{}{}
\setcitation{}{}
\setcitation{}{}
\setcitation{}{}
\setcitation{}{}
\setcitation{}{}
\setcitation{}{}
\setcitation{}{}
\setcitation{}{}
\setcitation{}{}
\setcitation{}{}
\setcitation{}{}
\setcitation{}{}
\setcitation{}{}
\setcitation{}{}
\setcitation{}{}
  
\makeatother
\begin{document}

\title{A web-based IDE for IDP}

\author{Ingmar Dasseville, Gerda Janssens}

\institute{KU Leuven}
\maketitle
\begin{abstract}
IDP is a knowledge base system based on first order logic. It is finding
its way to a larger public but is still facing practical challenges.
Adoption of new languages requires a newcomer-friendly way for users
to interact with it. Both an online presence to try to convince potential
users to download the system and offline availability to develop larger
applications are essential. We developed an IDE which can serve both
purposes through the use of web technology. It enables us to provide
the user with a modern IDE with relatively little effort.
\end{abstract}

\section{Introduction}

IDP\cite{WarrenBook/DeCatBBD14,TPLP/BruynoogheBBDDJLRDV} is a knowledge
base system based on an extension of first order logic: \fodot\cite{iclp/DeneckerV08}.
The goal is to split classical programs up into two parts: a declarative
part (the domain knowledge) and an imperative part (the tasks done
with this knowledge). Hence the name IDP: Imperative Declarative Programming. 

The language has proven promising in practical situations, but to
better support new and existing users, an IDE is an indispensable
tool for any new language. In order to facilitate this need, we developed
a web-based IDE for \fodot. The tool is available online at \url{http://dtai.cs.kuleuven.be/krr/idp-ide/}.
This new IDE supports the basic features you would expect from an
IDE such as syntax highlighting, error/warning visualization, and
auto-indentation. On top of this there are also a few features specific
to IDP. An example is the unsat core visualization, this helps the
user debug a program by indicating what part of a theory is inconsistent.
The editor has been used successfully in a course teaching IDP at
the KU Leuven, and the online version has already attracted new users
for IDP who use it in e.g. the tax administration business. 

IDP had one previous IDE\cite{url:idpide}. It was based on the Eclipse
Environment~\cite{url:PDE} and XText~\cite{url:xtext}, however
it was tedious to keep this IDE up to date with new Eclipse versions
and IDP language updates. In this new IDE we used a more lightweight
approach, while relying on the maturisation of web technologies. This
enabled us to provide more features with a smaller code base and to
host an online version of our IDE, which can be integrated into a
website to present the system. 

In the following sections we introduce the basics of IDP, give an
overview of the features of the IDE, give some insights in the implementation
of the IDE and finally compare our IDE with other IDEs for systems
that are similar to IDP.

\subsection{IDP}

IDP is a knowledge base system. IDP programs typically consist of
4 objects. A vocabulary, theory and structure are used to declaratively
define the knowledge which is relevant for the program. The procedure
then further explains what the system should do witht this knowledge.
\begin{description}
\item [{Vocabulary}] The vocabulary declares which types/predicates/functions
will be used.
\item [{Theory}] The theory is the representation of the knowledge over
a vocabulary, expressed in \fodot. It makes use of the standard first
order concepts such as $\forall,\exists,\Rightarrow,\wedge,\vee,\lnot$
extended with arithmetic, types, and inductive definitions. 
\item [{Structure}] A structure is a (partial) interpretation of the vocabulary.
It should at least interpret the types of the vocabulary and can further
interpret as little or as much of the rest of the vocabulary as desired.
Models of a theory fully interpret all symbols so that the theory
is satisfied. 
\item [{Procedure}] IDP uses a Lua scripting environment to interact with
the above objects. A typical procedure would be to print the models
of the theory, but this procedure could also include an interactive
simulation of a system which is described in the theory. 
\end{description}
IDP offers a lot of built-in procedures - inferences - which operate
on the above objects. Some examples of these are:
\begin{description}
\item [{Modelexpand(Theory,Structure)}] This inference produces a model
satisfying a theory and expanding a partial structure. The most typical
main procedure consists of printing the models returned by this method.
\item [{Unsatcore(Theory,Structure)}] This inference produces an unsat
core, which is a subset of the theory which is inconsistent with the
structure. This is useful for debugging purposes.
\item [{Propagate(Theory,Structure)}] Refines the structure with logical
consequences of the theory and the structure. 
\item [{Progress(Theory,Structure)}] Returns the set of possible next states
(structures) for a linear time calculus (time dependent) theory. This
can be used to simulate a dynamic system as explained in \cite{iclp/Bogaerts14}. 
\end{description}

\subsection{Approach to the IDE}

The IDP system is comparable to ASP systems, for which a multitude
of IDEs were developed, which are explored further in Section \ref{sec:Other-IDEs}.
Our approach differed from most of the others in the sense that we
tried to develop a more lightweight system. A lot of the IDEs try
to develop a lot of tools from scratch. However, good frameworks are
already available for developing IDE tools. We chose for a web-based
approach, based on existing frameworks. This approach provided us
with some advantages.

A web-based tool is agnostic about the operating system it runs on.
This means that no special code is needed differentiating the different
operating systems, while still providing a modern-looking layout.

Another advantage is that we can provide the IDE as a local application
as well as a website. The local application is based on a locally
run web server. The workspace folder and the IDP run command can be
set via a configuration file. The application can be accessed by browsing
to localhost. The web application is the same tool as the local application
with some added restrictions. The workspace folder is not writable
via the application and the resources for running files are limited.

\section{Overview of the IDE}

\begin{figure}
\centering\includegraphics[width=1\textwidth]{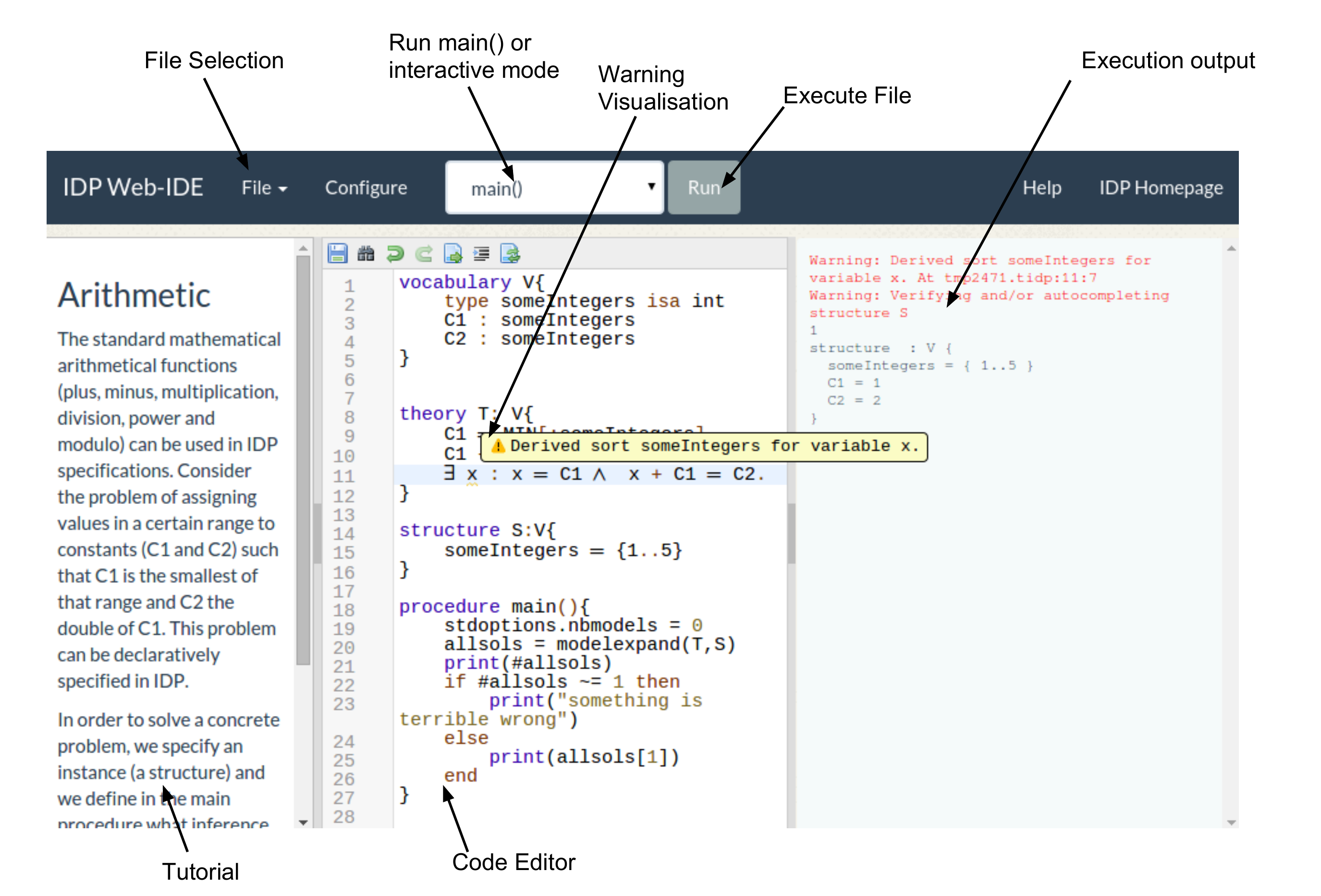}

\protect\caption{User Interface of the IDE}
\end{figure}

\subsection{Features}

\subsubsection{Editing}

Most of the time a user spends in an IDE, he is editing the source
code. It is the responsibility of the IDE to support the user as well
as possible. Apart from the obvious editing of text, a user expects
facilities such as highlighting errors and auto indentation.

\paragraph{Basic Text Editing}

The IDE provides all the features you would expect from a simple text
editor. This includes picking a file to edit, providing a text area
to edit the code, and supporting the most common general editing features
such as undo/redo, find/replace and line numbers.

\paragraph{Autocompletion/Code Snippets}

IDEs have often very different notions of autocompletion. Some IDEs
perform a full semantic analysis to suggest the possible correct completions
of the current word. The autocompletion in this IDE is more simplistic.
It is based on all the words occurring in the current file combined
with known code snippets. These snippets are predefined and consist
of a set of common pieces of code. These include declaration of components
such as theory and vocabulary, and for example a reachability relation.
Autocompletion can be triggered through the ctrl-space shortcut.

\paragraph{Syntax Highlighting}

Syntax highlighting is done on a syntactical basis. The theory is
divided into tokens, they indicate the function of the word in the
source coding, such as logical symbol, comment or keyword. This information
is then used to color certain parts of the code.

\paragraph{Logical Symbol Replacement}

IDP makes heavy use of ASCII representations of mathematical symbols.
The tokenized version of the theory identifies the logical symbols.
The editor automatically replaces these symbols with their mathematical
counterparts.

\paragraph{Auto indentation}

Uniform indentation vastly increases the readability of files. The
editor supports the user to use a consistent indentation and provides
the possibility to reindent the whole document.

\paragraph{Error/Warning visualization}

Syntax errors and warnings are visualized right where they occur through
zig-zag lines. Without this feature, users spend a lot of time tracking
down the exact location of an error. This feature is implemented by
running the internal parser of IDP, so the feature always uses the
current version of the syntax. This prevents the IDE from being unusable
when a new language construct is added to the language.

\paragraph{Code Sharing}

The editor (both local and server) provides the capability to export
code and provide a global link for sharing the code. This part of
the editor uses a GitHub Gist-account\cite{url:gist} to back-up the
code.

\subsubsection{Running}

Apart from editing, the IDE can also provide enhanced support for
running the files a user wrote. The most basic support is simple terminal
support. But this IDE also supports some extra visualization features.

\paragraph{File Running}

\begin{figure}
\centering\includegraphics[width=1\textwidth]{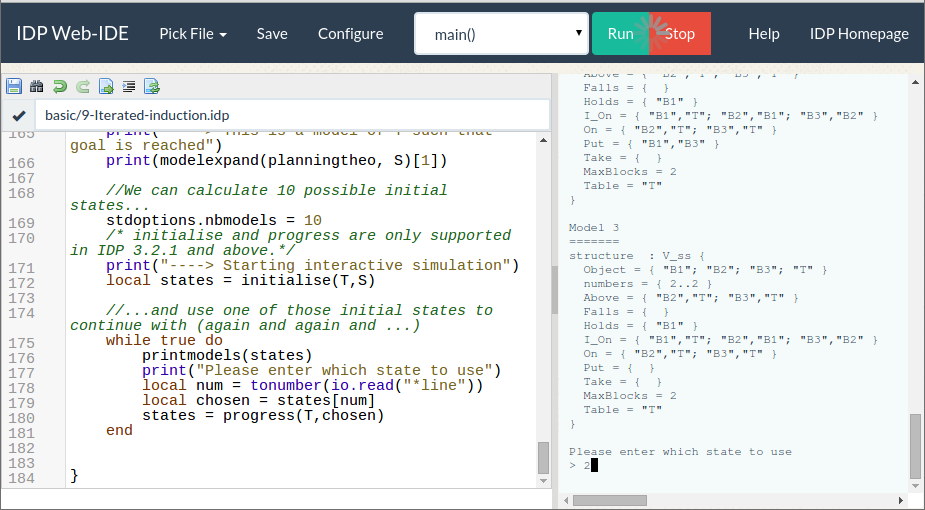}

\protect\caption{Interactive Running\label{fig:Interactive-Running}}
\end{figure}

Executing a file is supported in the IDE. This includes the possibility
of an interactive session, or a program where a Lua procedure needs
to ask input from the user as can be seen in Figure \ref{fig:Interactive-Running}. 

\paragraph{Lua Shell Mode}

Apart from running the main procedure of a specific file, it is also
possible to select ``Interactive Mode'' in the dropdown menu to
the left of the run button. This starts an interactive Lua Shell so
the user can interact directly with the components in the file.

\paragraph{Unsat core visualization}

\begin{figure}
\centering\includegraphics[width=0.8\textwidth]{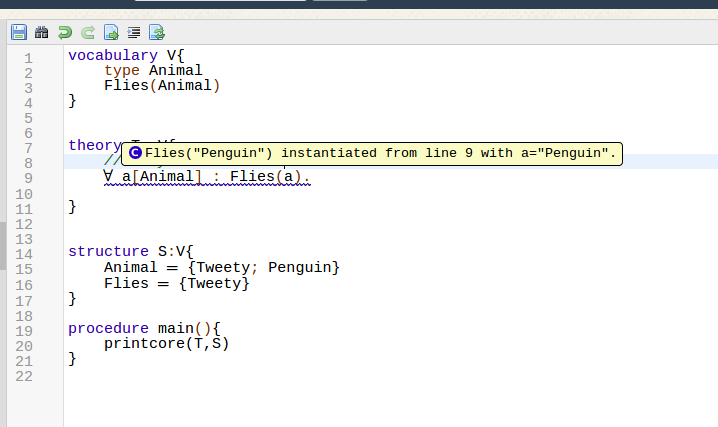}

\protect\caption{Unsatcore Visualization\label{fig:Unsatcore-Visualisation}}
\end{figure}

IDP supports the user throughout the debugging process of unsatisfiable
theories with the help of unsat core extraction\cite{kbse/ShlyakhterSJST03}.
This process extracts a subset-minimal part of the theory which is
inconsistent over a given structure. The result is thus a set of ground
formulas which are instantiations of lines in the theory. This information
can be visualized in the editor with zig-zag lines at the appropriate
theory lines together with a tooltip indicating the instantiations
in the unsat core. This visualization can be seen in Figure \ref{fig:Unsatcore-Visualisation}.
In this example, the structure specifies there are 2 animals, of which
one can fly. This is not consistent with the theory stating that all
animals fly. The core explains the unsatisfiability of the theory
with the instantiantion of the sentence where the unsatisfying animal
is the penguin.

\paragraph{Visualizations}

\begin{figure}
\centering\includegraphics[width=1\textwidth]{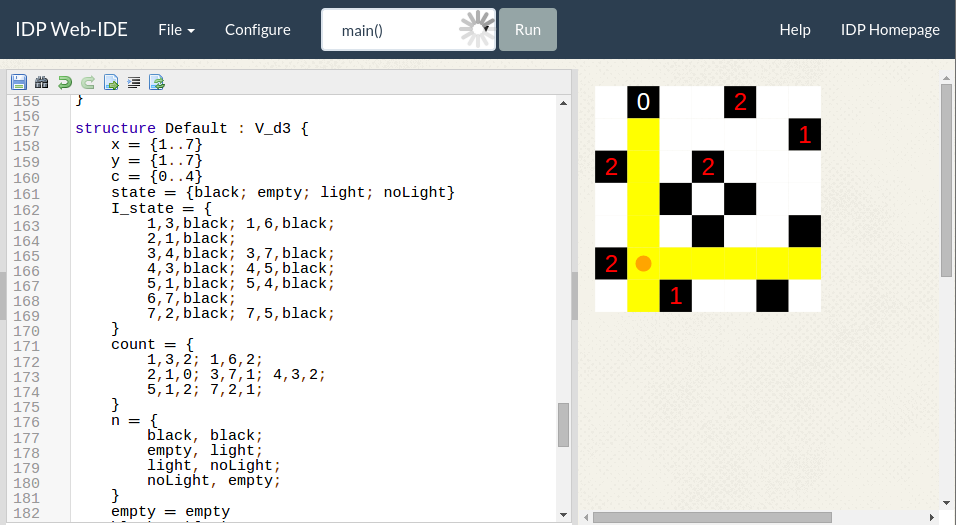}

\protect\caption{Visualization of an interactive Lights Out puzzle\label{fig:IDPD3}}
\end{figure}

IDP provides the possibility to write theories over a graphical vocabulary
so that the output can be visualized. This visualization can even
be interactive, so it can react to clicks of the user. This visualization
technology is called IDPD3\cite{lapauw2015idpd3}, it uses the same
philosophy as ASPVIZ\cite{cliffe2008aspviz} and IDPDRAW\cite{url:idpdraw}
extended with interaction. This technology is mostly used to visualize
the output of a search problem but can also be used to provide an
interactive puzzle to a user, as can be seen in Figure \ref{fig:IDPD3}.
When a visualization method is called, a pane with the image is automatically
opened. 

\subsubsection{Online-Specific Features}

\begin{figure}
\centering\includegraphics[width=1\textwidth]{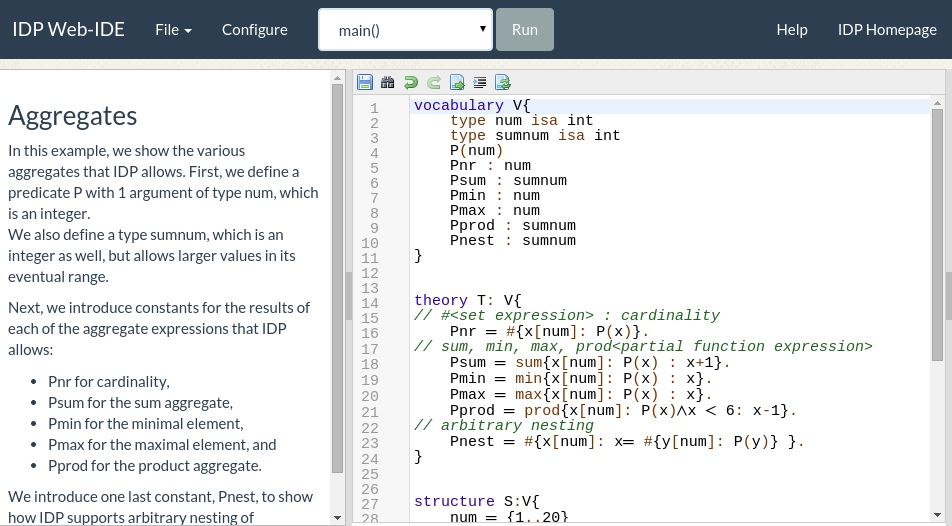}

\protect\caption{Tutorial mode of the editor\label{fig:tut}}
\end{figure}

A few modifications are used for the online version of the IDE. In
the online version of the IDE, a pane with explanations can be automatically
coupled to an example. In this way, a tutorial with examples can be
completely integrated with the IDE. So, the IDE can be used as a learning
environment. The public website of the IDE is structured in this way,
an example of this can be seen in Figure \ref{fig:tut}. 

The security aspect enforces a restriction of some of the features.
When providing a system online, precautions need to be taken so that
the system is safe. In a standard setting, the Lua environment of
IDP provides full access to the host system. As a consequence, unsafe
methods should be disabled. Also resource limitations are imposed
upon requests coming from the web.

\subsection{Implementation}

\begin{figure}
\centering\includegraphics[width=1\textwidth]{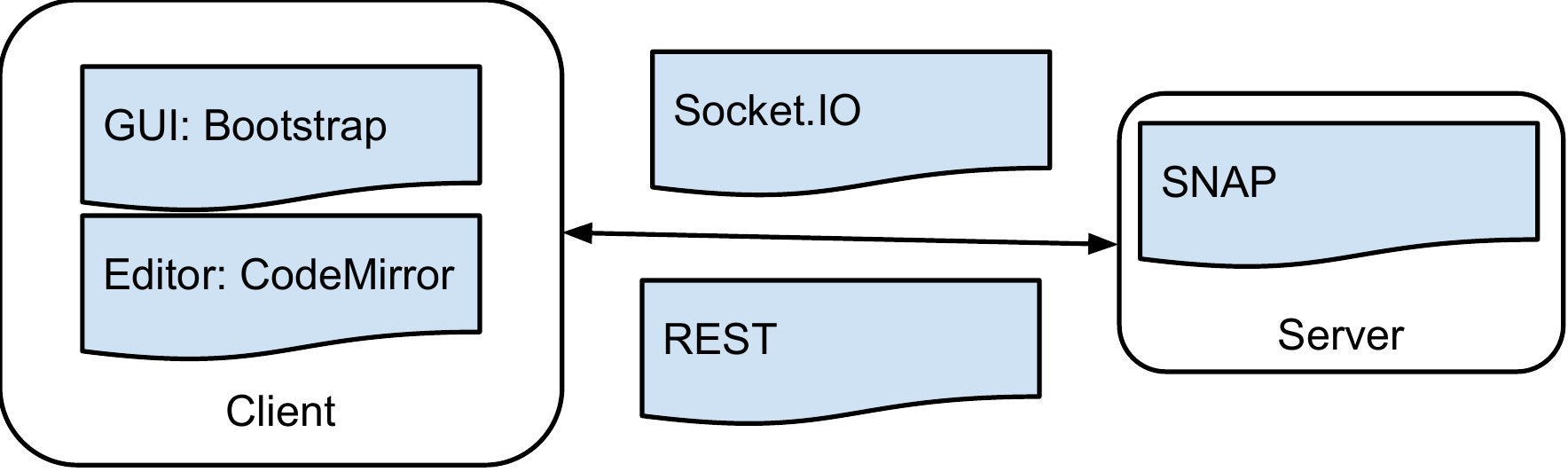}

\protect\caption{Overview of the IDE architecture\label{fig:Arch}}
\end{figure}

The client side of the IDE completely consists of JavaScript, CSS
and HTML code and the server is written in Haskell. Libraries were
used for the generic components of the UI. The general structure of
the IDE can be seen in Figure \ref{fig:Arch}.

\begin{description}
\item [{CodeMirror}] CodeMirror\cite{url:codemirror} is a JavaScript-based
text editor with an API which has good support for extensions and
new languages. All text-editing features are implemented on top of
this editor. Features such as syntax highlighting are natively supported
and only require simple, language-specific code to be added. 
\item [{Bootstrap}] This state-of-the-art CSS/JavaScript framework developed
by Twitter~\cite{url:Bootstrap} allows for the quick creation of
a simple and effective user interface. It is well supported by all
browsers, so we can ensure that the IDE works equally well on all
operating systems. 
\item [{Snap}] To communicate between the CodeMirror instance in the web
browser and IDP, we wrapped IDP into a RESTful interface. This is
the most common technique on the web for exposing an API. This interface
is written in Haskell using the Snap\cite{url:snap} framework.
\item [{Socket.io}] When running an interactive IDP session, we needed
to keep a session open during its runtime. To support this communication,
we chose for WebSockets. This is a fairly new technology, but luckily
socket.IO\cite{url:socket.io} provides an automatic fallback mechanism
whenever a browser does not support WebSockets. In this case, the
application will run through traditional XHR/JSONP requests which
simulate the WebSockets.
\end{description}
With the use of these technologies, the actual code of the IDE can
be very compact, e.g. the web server consists of less than 300 lines.
We also expect that future changes to the IDP language require very
little change to the IDE as features such as the error reporting are
fully dependent on the parser of the IDP executable. This approach
to parsing also has disadvantages. It is difficult to implement advanced
features such as smarter autocompletion without having a full parser
available in the IDE itself. We judged that for this editor the costs
did not outweigh the benefits.

\section{Other IDEs\label{sec:Other-IDEs}}

Existing editors for truly-declarative languages such as \fodot and
ASP can be roughly divided in two groups: custom IDEs which are built
from the ground up and Eclipse plugins. This IDE falls into a third
category, the web-based IDEs. As far as the authors know, such editors
only exist for imperative languages.

\subsection{Custom IDEs}

Custom IDEs often lack some basic features such as find/replace or
undo/redo, which vastly undermines their practical usability. An example
of such an editor is OntoDLV\cite{ricca2009ontodlv}, which does not
support undo. ASPIDE\cite{febbraro2011aspide} seems to be the most
comprehensive custom IDE which does include basic features such as
find and undo. It has an impressive feature list, including quick
fixes for code. However, the use of Java GUI tools causes the GUI
to be platform independent. This is easy from the programmer point
of view, but OS-specific things that users expect in all applications,
such as shortcuts, are not there.

\subsection{Eclipse Plugins}

Eclipse plugins benefit from the good integration Eclipse provides
for new languages. However, there are some disadvantages to this approach.
Eclipse requires a complete new parser for the language. This means
that the IDE needs an update at every syntax change of the language.
It is sometimes difficult to ensure the real language and the Eclipse
language consist of the same grammar, thus it occurs often that either
the parser is not strict enough and some syntax errors go unnoticed
until the file is run, or some errors are indicated which are actually
acceptable constructs.

Examples of ASP editors in this category are Videas\cite{oetsch2011videas}
and SeaLion\cite{oetsch2013sealion} with the visualization tool Kara
\cite{kloimullner2013kara}. SeaLion seems like the most modern ASP
IDE. It has a tight integration with Eclipse, and provides features
such as a debugger and visualizations.

\subsection{Web-based IDEs}

Web-based IDEs are a relatively recent phenomenon, but have gained
some traction in the last few years. Cloud9\cite{url:cloud9} is one
of the more popular ones. It provides the user with all tools he uses
during development of an application, including full access to a virtual
machine. The IDP web-IDE belongs to this category. 

Our IDE has one advantage that most of the others have not. It is
one of the few tools which support both offline and online use. Two
other (general purpose) IDEs that support this are Codiad~\cite{url:Codiad}
and ICEcoder~\cite{url:ICEcoder}. Both tools use PHP for their back-end
and ace-editor for front-end editing. Their editing features are more
extensive than our IDE, especially the support for folders and projects.
However, there is no integration for running the files which are being
edited. 

Some ASP systems do have an online tool available. The DLV system
has an online version since 2006 available at \url{http://asptut.gibbi.com/}.
This website mainly consist of a plain textbox and a run button. It
mainly serves the purpose to follow a tutorial in the form of slides.
Clingo has a more recent tool available at \url{http://potassco.sourceforge.net/clingo.html}
including syntax highlighting.

One system in this category is special. Atom~\cite{url:atom} is
a purely offline editor which is also based on web technology. It
essentially consists of a web browser with less security restrictions,
so it integrates better with the local filesystem, but there is no
way to run Atom as a web service.

\section{Integration in other web applications\label{sec:Integration-in-other}}

An advantage of our approach is that it makes it possible to include
parts of the IDE into other web-based applications. Inserting an IDE
in an application is as simple as loading the CodeMirror library with
the right set of options. Running a file can be done through the REST
interface which is exposed by an IDE server.

We currently have 2 demos online which make use of this approach.
The first one demonstrates the selection of courses in a study programme.
Through the inclusion of the editor we can provide the possibility
for the user to interactively modify the underlying theory of this
course selection system. This demo is available at \url{http://krr.bitbucket.org/courses/}.
The second one demonstrates a problem where PC's should be configured
to have the right software, this demo is available at \url{http://krr.bitbucket.org/desired/}.

\section{Conclusion}

In this paper we presented the IDP Web-IDE. This new IDE for the IDP
system was developed using web technology. This enabled us to provide
an online as well as an offline version of the editor. The IDE supports
most of the basic features you would expect from an IDE. We believe
that the web-based approach is a new setting for IDEs and that this
setting works inspiring e.g. to add new features. The recent addition
of the tutorial mode is such an example. 

However, some features from other IDEs, such as managing multiple
files as a project are currently not supported. We intend to expand
the the IDE so it can continue to serve as a teaching and development
tool. We also want to make the tools in the IDE available for other
users so it becomes easier for other developers to make tools like
in Section \ref{sec:Integration-in-other}.

The IDE is publicly available at \url{http://dtai.cs.kuleuven.be/krr/idp-ide/}.
There, the server version can be tried and a link to download the
IDE for offline use is provided.

\bibliographystyle{plain}
\bibliography{idp-latex/krrlib,resources/ideresources}

\end{document}